\documentclass[twocolumn,trackchanges]{aastex631}

\usepackage{graphicx}
\usepackage{xcolor}

\usepackage{xspace}
\newcommand{\Lya}{Ly$\rm \alpha$\xspace}

\submitjournal{ApJ}

\shorttitle{Low $f_{\rm esc}(\rm{LyC})$ in three Massive, High-z Galaxies}
\shortauthors{Witten, Laporte and Katz}

\graphicspath{{./}{figures/}}

\begin{document}

\title{Evidence for a low Lyman Continuum Escape fraction in three Massive, UV-bright galaxies at $z > 7$}

\correspondingauthor{Callum E. C. Witten}
\email{cw795@cam.ac.uk}

\author[0000-0002-1369-6452]{Callum E. C. Witten}
\affiliation{Institute of Astronomy,
University of Cambridge,
Madingley Road, Cambridge CB3 0HA, UK}

\author[0000-0001-7459-6335]{Nicolas Laporte}
\affiliation{Kavli Institute for Cosmology,
University of Cambridge,
Madingley Road, Cambridge CB3 0HA, UK}
\affiliation{Cavendish Laboratory,
University of Cambridge,
19 JJ Thomson Avenue, Cambridge CB3 0HE, UK}

\author{Harley Katz}
\affiliation{Sub-department of Astrophysics, University of Oxford, Keble Road, Oxford OX1 3RH, UK}

\begin{abstract}
Although low-mass star-forming galaxies are the leading candidates of the reionisation process, we cannot conclusively rule out high-mass star-forming galaxies as candidates. While most simulations indicate the former is the best candidate some models suggest that at $z \geq6$ massive, UV-bright galaxies - ``oligarchs" - account for at least 80$\%$ of the ionising budget. To test this hypothesis we target massive ($\log_{10} \rm (M_{\star} [M_{\odot}]) >10$), UV-bright ($M_{UV} \sim-22$) \Lya emitters at $z >7$ in archival data, observed with similar resolution spectrographs (VLT/X-shooter and Keck/MOSFIRE). To increase the reliability of our conclusions we stack all spectra and obtain a deep-stacked spectrum of 24.75hrs. The stacked Ly-$\alpha$ profile displays a clear asymmetric red peak and an absence of a blue peak. We additionally estimate the intrinsic stacked \Lya profile of our targets by correcting for IGM transmission using a range of neutral hydrogen fractions, finding no significant change in the profile. We measure a velocity offset V$_{\rm red} > 300$ km/s and an asymmetry in our red peak $A \sim 3$. Using various models and estimators such as the peak separation, the asymmetry of the red peak, the ratio between \Lya and $\rm{H}\beta$ and the $\beta$ slope, we conclude that the  escape fraction in these three UV bright, massive ($\sim 10^{10} M_{\odot}$), $z\geq7$ galaxies is $f_{\rm esc}\rm{(LyC)}\leq10$\%.

\end{abstract}

\keywords{}

\section{Introduction} \label{sec:intro}
Cosmic reionisation is the processes by which the neutral hydrogen in the universe, formed after the recombination phase (380 000 years after the Big-Bang), underwent a phase change - to almost complete ionisation by $z = 5.5$ (\citealt{Bosman2022}, \citealt{Kulkarni2019}, \citealt{Robertson2015}). Although the period over which reionisation occurs is well constrained, the sources of the ionising photons are not. The current debate is regarding whether or not the faint galaxies which represent the bulk of galaxies at $z\geq6$ (e.g. \citealt{Bouwens2015}, \citealt{Finkelstein2015}, \citealt{Atek2018}) are the main drivers (\citealt{Ocvirk2020}, \citealt{Trebitsch2022}) or if the most massive objects are the major contributors \citep{Naidu+20, Sharma+17} to reionisation. For example, \citet{Naidu+20} claim that bright galaxies may produce $> 80 \%$ of the reionisation budget in order to match the observed rapid fall in neutral hydrogen fraction at $z < 8$. Directly measuring the Lyman continuum radiation escaping a source and hence ionising the universe is however not possible due to the neutrality of the IGM during the epoch of reionisation \citep[eg.]{Inoue+14}. Instead the Lyman-$\alpha$ profile is heavily affected by neutral hydrogen and hence it can be used as a probe of the amount of ionising photons escaping the galaxy. 

Given the potential \Lya emission has as a probe of neutral hydrogen, research into the properties of the line emerging from a source through surrounding neutral hydrogen were conducted by \citet{Harrington+73, Neufeld+90}. The absorption of \Lya at line centre by neutral hydrogen results in the main escape mechanism for photons being through diffuse in frequency producing a double-peaked spectrum. The wings of the \Lya line, that form the consequent two peaks are defined by the density of the scattering neutral hydrogen. As such the separation between the two peaks ($V_{\rm sep}$) has been proposed as a key diagnostic of the Lyman-continuum escape fraction $f_{\rm esc}$(LyC) due to their tight dependence \citep{Izotov+17, Verhamme+15}.

While this double-peaked spectrum has now been observed \citep[eg.][]{Yee+91, Venemans+05, Vanzella+08}, especially at high-redshift, we often see reduced or a complete absence of the blue peak \citep{Verhamme+15}. This process is now well understood as a result of inter-galactic medium (IGM) radiative transfer (RT) effects \citep{Gunn+65} at high redshifts where we do not expect to find strong outflows (\citealt{Vito2022}, \citealt{2005ApJ...618..569M}). Photons are redshifted as they travel through the expanding universe, and hence blue peak photons get shifted into resonance. If this occurs in the vicinity of neutral hydrogen, we observe absorption, and hence any intrinsic blue peak is not observed for high neutral hydrogen column densities as predicted by \citet{Garel+21, Laursen+11}. 

Emission from high redshift objects typically travels through high neutral hydrogen patches in the IGM and as such, we rarely observe blue \Lya peaks. There have however been instances of double-peaked high redshift galaxies such as that discussed in \citet{Meyer+21} and \citet{Matthee+18}. These galaxies typically live within large ionised bubbles resulting in a largely reduced neutral hydrogen column density, allowing the presence of a blue-peak. 

In this letter, we aim to measure the mean LyC escape fraction of bright  $z\geq$7 Lyman Alpha Emitters (LAEs) based on their \Lya profile and to give new insights on their contribution to the reionisation. In section~\ref{sec:Targets}, we describe the sample of LAEs we identified at $z\geq$7. We then describe the stacking method we use to improve the signal-to-noise ratio on our \Lya line in section~\ref{sec:Method} and we discuss the results and their implications in sections~\ref{sec:Results} and~\ref{sec:Discussion}.

\section{Targets selection} \label{sec:Targets}
\begin{table*}[t]
\caption{The properties of the archival observations of our three target galaxies. $z_{syst}$ indicates the systemic redshift of the target galaxy that we use for our analysis and the stellar mass is determined by the BAGPIPES fitting discussed in Section~\ref{sec:Targets}.}

\label{tab:Targets}
 \begin{tabular}{lcccccc}
  \hline
  Name & $z_{syst}$ & log($M_{\star}$) & Exposure time & \Lya luminosity  & Telescope & Reference \\
       &            & $[M_{\odot}]$ &  & [10$^{43}$ erg/s] & & \\
  \hline
  EGSY-8p68 & 8.671 & $10.1^{+0.1}_{-0.2}$ & 4 hrs. 45 min. & 1.95±0.49 & MOSFIRE & \citet{Zitrin+15}\\
  EGS-zs8-1 & 7.721 & $10.2^{+0.2}_{-0.1}$ & 4 hrs. & 1.2±0.1 & MOSFIRE & \citet{Tilvi+20}\\
   & & & 4 hrs. & 1.2±0.2 & MOSFIRE & \citet{Oesch+15}\\
  COSY & 7.142 & $10.2^{+0.1}_{-0.8}$ & 12 hrs. & 1.43±0.19 & X-Shooter & \citet{Laporte+17}\\
  \hline
 \end{tabular}
\end{table*}
Our primary goal is to infer the mean escape fraction of bright, $z\geq$ 7 galaxies by constraining the shape and properties of \Lya, 1215.67\AA , redshifted into the near-infrared. We therefore search in the literature for all $z\geq$7 spectroscopically confirmed galaxies that have been observed by high-resolution NIR spectrographs, to guarantee the ability to resolve any double-peak in the \Lya profile. Moreover, to obtain a good estimate of the UV beta slope, which can be used to measure the escape fraction at high-redshifts \citep{Zackrisson+17}, we require that our selected galaxies have at least 3 constraints on their SEDs. 11 objects with a redshift ranging from 7.15 to 9.11 were selected with \Lya luminosities ranging from $\sim$0.05 to 2$\times$10$^{43}$erg/s.

Additionally, \citet{Schenker+12} reports the detection of A1703-zd6, a redshift $z = 7.045$ galaxy that satisfies all of our selection criteria, except one - it is a lensed, low-mass galaxy (\citealt{Stark+15} - $\log$($M_{\star}) \sim 8.7$). We therefore do not include this target within our stack, but we note the measured \Lya red-peak separation from line-center of $\sim$ 60 km/s. This is consistent with a high Lyman-continuum escape fraction as expected if faint galaxies are responsible for re-ionization \citep{Gazagnes+20, Izotov+18a}.

Because the observed shape of the \Lya line profile is a result of propagation through neutral gas, it does not trace the precise redshift of a galaxy. In order to produce a proper stack of this emission line, we also require that the galaxies used in our study have a systemic redshift measurement through the observation of another emission line. Among the 11 galaxies mentioned above, only  3 (described in table~\ref{tab:Targets}) satisfy this criteria, returning 4 datasets combining 24 hours and 45 minutes on-source exposure time. For MOSFIRE, we make use of the following programs : C228M (PI: A. Zitrin), Y288M (PI: Moncheva) and N190 (PI: Malhotra) ; and for XSHOOTER : 097.A-0043(A) (PI: Ellis). The three aforementioned galaxies have UV luminosities $M_{UV} \sim -22$, placing them on the extreme end of galaxies luminosities at $z = 7-8$ \citep{Bowler+14, Bowler+20}. 

The Spectral Energy Distribution (SED) of the 3 galaxies have been extracted from 3DHST catalogues (\citealt{Brammer2012}, \citealt{Skelton2014}). The physical properties of each individual galaxy have been estimated by SED-fitting using BAGPIPES \citep{BAGPIPES}. One of the main advantages of this code is to allow the user to choose between several Star Formation Histories (SFHs). In this work, we run BAGPIPES with 4 different SFHs, namely a burst, a constant, a delayed and a combination of burst+constant. The best SED-fit is obtained for the SFH that minimised the BIC (see \citealt{Laporte2021b} for more details). BAGPIPES being a parametric code, we used the following ranges for the parameters of the SFH : 
\begin{itemize}
\addtolength\itemsep{-2mm}
\item ionisation parameter : $\log$ U $\in$ [-3.0,-1.0]
\item dust attenuation (assuming a Calzetti law) : $A_v[mag] \in$ [0.0, 1.0]
\item age of the stellar population : Age[Gyr] $\in$ [0.0,1.0]
\item mass formed : $M_{\star} [M_{\odot}] \in$ [$10^6$, $10^{12}$]
\item metallicity : Z [$Z_{\odot}$] $\in$ [0.0, 1.5] 
\end{itemize}

The redshift was fixed to the spectroscopic redshift of each galaxy, and the IMF used in BAGPIPES is a Kroupa IMF \citep{KroupaIMF}. As expected by the selection function of our sample, our galaxies are good examples of the most massive galaxies at $z\geq$7 with stellar masses ranging from 1.36$\times$10$^{10}$ to 1.75$\times$10$^{10}$ $M_{\odot}$, placing them on the extreme end of the galactic stellar mass function at $z = 7-8$ (see Table~\ref{tab:Targets}). The properties of our stacked spectrum obtained using BAGPIPES are found in Table~\ref{tab:Targets}.

The detection of \Lya at $z\geq$7 implies the formation of an ionized bubble around the galaxies. \citep[e.g.][]{Castellano2022, Roberts-Borsani2022}. The origin of the ionising photons that produce these ionised bubbles is still highly debated and could be either due to the intrinsic nature of the object (e.g. star formation or an active galactic nucleus) or the over-dense environment near the most massive galaxies formed at high-redshift \citep{Leonova+21, Laporte+22}. Indeed previous research into these galaxies indicates that they reside within ionised bubbles,  large enough that any blue-peak escaping the host galaxy would be redshifted past \Lya line-centre before leaving the ionized bubble and hence should be unaffected by IGM absorption. 

Based on a relation between \Lya luminosity and ionized bubble size derived from theoretical models, \citet{Tilvi+20} estimate that EGS-zs8-1 sits in a common bubble with at least 3 neighbouring galaxies, with a size of 1.02 pMpc. More recently, \citet{Leonova+21} present analysis of \textit{Hubble} Space Telescope (HST) imaging that supports the conclusion of a bubble surrounding EGS-zs8-1 given it resides in an overdensity. They additionally find that EGSY-z8p68 is found with an overdensity and again comparing to simulations find an expected bubble radius of $\sim 1$ pMpc. \citet{Laporte+17} detect \Lya, HeII and NV, as well as upper bounds on the flux of CIII] and CIV in the spectrum of COSY allowing for the determination of the source of its radiation field. The most likely hypothesis is that the radiation field of COSY is inconsistent with that from star-forming galaxies, instead it likely is produced by an active galactic nucleus (AGN) \citep{Laporte+17, Costa+14} and additionally its high UV luminosity makes it likely to trace an overdense region \citep[eg.][]{Barkana+04,Furlanetto+04}. Therefore it is likely COSY additionally resides within a large, ionized bubble.

\section{Method} \label{sec:Method}

\begin{table*}[t]
\caption{The values for various parameters used for a range of diagnostics, and the $f_{\rm esc}(\rm{Ly\alpha})$ that these diagnostics predict.}

\label{tab:fesc}
 \begin{tabular}{lcc}
  \hline
  Diagnostic & Value & $f_{\rm esc}(\rm{Ly\alpha})$ \\
  \hline
  Spitzer: flux (H$\beta$ + [OIII]) & $7.13 \times 10^{-17} \rm{erg/s/cm^{2}/}$\r{A}  & [0.02:0.32]\\
  BAGPIPES: flux (H$\beta$) & $1.1^{+0.9}_{-0.5} \times 10^{-17} \rm{erg/s/cm^{2}/}$\r{A} & $\sim$ [0.09:0.18]\\
  Spitzer: UV slope $\beta$ and log$_{10}$(EW(H$\beta$ + [OIII])) & [-0.6, -2.35], [2.0 \r{A}, 3.1 \r{A} ] & $\sim 0$\\
  BAGPIPES: UV slope $\beta$ and log$_{10}$(EW(H$\beta$)) & $-1.89^{+0.09}_{-0.11}$, $2.67 \pm 0.25$ & $\sim 0$\\
  \Lya profile: Asymmetry and Peak separation & $A = 3.3^{+1.4}_{-0.8}$, $V_{\rm{red peak}} = 330^{+190}_{-70}$ km/s & $< 0.15$\\
  
  \hline
 \end{tabular}
\end{table*}

Data reduction of archival Keck MOSFIRE and VLT XSHOOTER data was performed through the standard MOSFIRE data reduction pipeline\footnote{\url{https://keck-datareductionpipelines.github.io/MosfireDRP/}} and EsoReflex\footnote{\url{https://www.eso.org/sci/software/esoreflex/}} which include standard data reduction procedures such as flat-fielding, wavelength calibration and background subtraction. The flux calibration was performed using a bright photometric standard observed during each night. 

The systemic redshift is obtained from the additional emission lines present in each galaxy's spectrum (EGSY-8p69: N V, \citet{Mainali+18}; EGS-zs8-1: C III], \citet{Stark+17}; COSY: [C II], \citet{Pentericci+16}). We then shift each spectrum into the rest frame wavelength based on the systemic redshift reported in Table~\ref{tab:Targets}. Following this we define a new wavelength basis, spanning the range of wavelengths of interest with a wavelength bin equal to that of our lowest resolution spectra ($\sim 0.15$ \r{A}). The total flux 
of each object within each wavelength bin is calculated, converted into a luminosity, and the median of these luminosities is then taken as the value for our stacked spectrum. We then return this median spectrum to units of flux by dividing through by the luminosity distance of the median redshift in our sample. We take the standard deviation of the fluxes in each bin to obtain the error in our stacked spectrum.

We then use a Monte Carlo (MC) error propagation method to estimate the uncertainties on the observed red-peak velocity offset and asymmetry. We use the error in our spectrum, described above, to redraw each spectral bin's flux from a Gaussian centred on the median stack value with a standard deviation equal to the aforementioned error. We then repeat this process, measuring the red-peak offset and asymmetry, one hundred thousand times in order to understand the uncertainty in our measurements driven by the error associated with each flux measurement. We take the red-peak offset and asymmetry associated with our stack and measure the standard deviation of those values above and below the median for our upper and lower bound uncertainties on this measurement respectively.

We have additionally evaluated the results when normalising all of our spectra by dividing through by their peak \Lya flux, as not to weight our stack based on the most luminous galaxy. Given that every galaxy has a similar luminosity, we see no change in our \Lya profile regardless of normalisation. A median stacking method is chosen as it acts to reduce the affect of sky-lines, however, in the case of COSY, significant contamination from a sky-line in a region of the spectrum where we expect to observe the tail of our red-peak profile did act to pollute our stacked spectrum and it was therefore masked. Adopting a mean stacking method and additionally changing the position and widths of bins provides a very similar stacked spectrum thus indicating the robustness of our stack. 

\section{Results} \label{sec:Results}
\begin{figure*}
	\includegraphics[width=\textwidth]{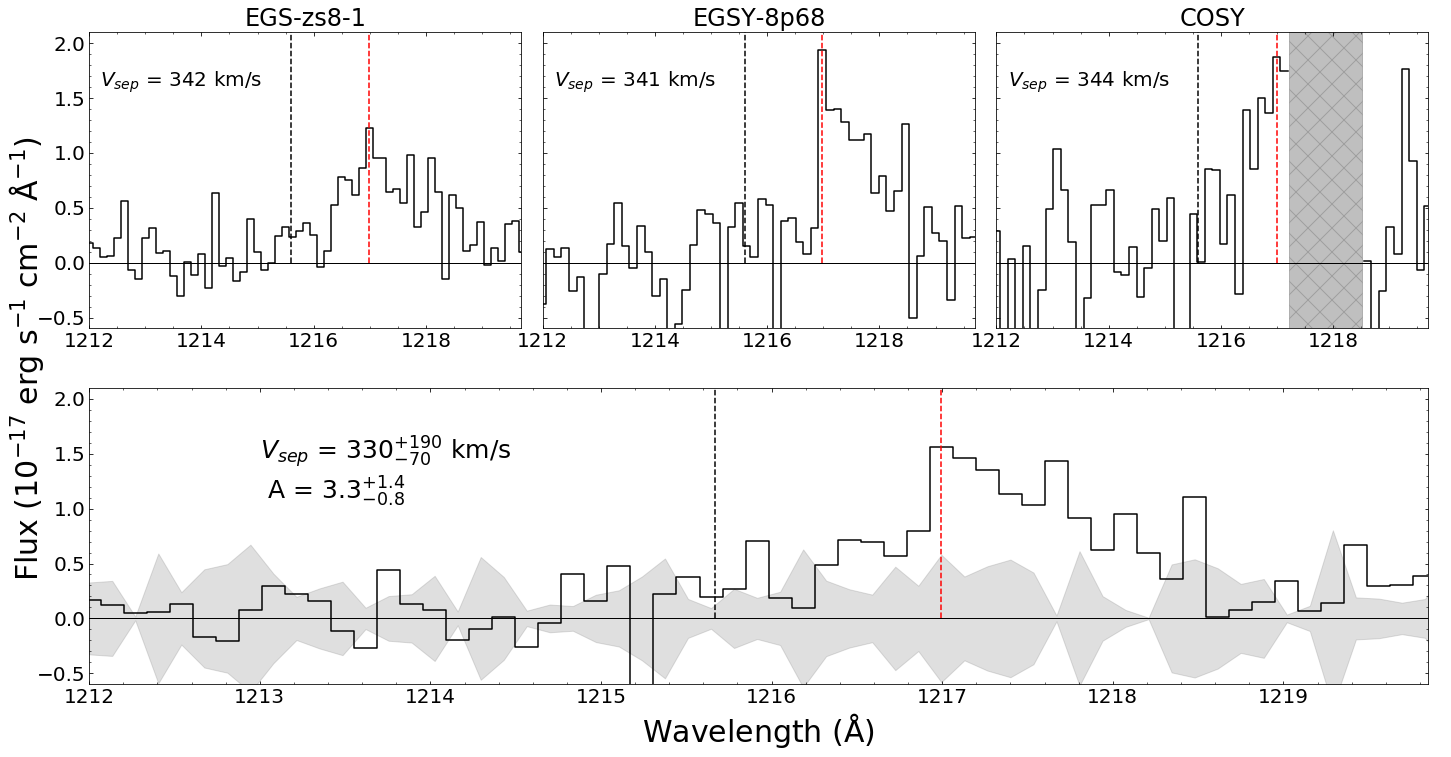}
    \caption{Line profiles of \Lya emission in our target galaxies (top) and the stacked spectrum (bottom). Line centre is denoted by a black dashed line, while the red-peak flux is indicated by a red dashed line. The velocity offset of the red-peak from line centre is additionally found within each panel. (Top:) From left to right: EGS-zs8-1, EGSY-8p68, COSY. The resolution of the spectrum of COSY has been reduced to the resolution of the two MOSFIRE spectra (0.14 \r{A}). The hatched box indicates a region of the spectrum that has been removed due to significant pollution by a sky line. (Bottom:) The grey region indicates the 1-sigma error obtained by taking the standard deviation of the constituent galaxies of the stack. Additionally, red peak velocity offset and the asymmetry of the red-peak as defined in Section~\ref{sec:Results} are found within the panel.}
    \label{fig:LyaProfile}
\end{figure*}
\subsection{Constraints from SED}
\label{sec:Photometry Results}

The \Lya escape fraction is the ratio of the \Lya flux escaping a galaxy over the intrinsic \Lya flux of the object. In order to ascertain limits on the potential intrinsic flux of \Lya we use recombination lines with known relations to \Lya. Unfortunately no direct observations of recombination lines are made in any of our target spectra. Instead we make use of two independent methods to estimate the equivalent width (EW) of the $\rm{H}\beta$ emission line: (i) using the flux ratio between the 4.5$\mu$m and 3.6$\mu$m assuming that the 3.6$\mu$m flux is the stellar continuum and the 4.5$\mu$m is the sum of the stellar continuum with a contamination of OIII+$\rm{H}\beta$ and (ii) using BAGPIPES to directly predict the flux of $\rm{H}\beta$ (see Table~\ref{tab:fesc} for the results of both methods). The first method returns the flux of $\rm{H}\beta$ and OIII in combination and thus by assuming no contribution from OIII supplies us with a lower bound estimate of the flux of the recombination line H$\beta$. We assume a maximum contribution of $\rm \log_{10} ([OIII]/\rm{H}\beta) = 1$ - based on the assumption our stack lies in the extreme AGN region of the Baldwin, Philips and Terlevich (BPT) diagram \citep{Baldwin+81, Veilleux+87}. This conclusion is in itself unlikely as NV emission lines are either weak or not present in our individual galaxy spectra but allows us to obtain an absolute upper bound on the $\rm{H}\beta$ flux. The second method is based on SED-fitting and depends on the best fit parameters such as the stellar mass, the reddening, the age, the metallicity, etc. To verify that the $\rm{H}\beta$ flux estimated by BAGPIPES is not strongly dependant on other parameters, we study the evolution of $\rm{H}\beta$ flux as a function of the metallicity and stellar mass and as a function of the metallicity and reddening. The variation in the EW is estimated as $\Delta \log{EW} <0.5$.

We assume, following \citet{Gazagnes+20}, that $\rm L_{Ly\alpha} = 8.7 L_{H_{\alpha}}$, and in turn that $\rm L_{H_{\alpha}} = 2.85 L_{H_{\beta}}$, however Ly$\alpha$ emission can include a large contribution from collisional excitation \citep{Mitchell+21, Smith+21} thus increasing the intrinsic $\rm L_{Ly\alpha}$ further decreasing the measured escape fraction. 

However, given we observe no \Lya blue-peak we must assume the \Lya profile has undergone some absorption due to IGM attenuation. Given many \Lya profiles of low redshift LAEs, that have not travelled through a high neutral hydrogen density IGM, exhibit often equal or less than equal blue-to-red peak flux ratios \citep[eg.][]{Izotov+18a,Izotov+18b} we assume that the upper bound of the $\rm L_{Ly\alpha}$ escaping the galaxy is double the $\rm L_{Ly\alpha}$ that we observe, while the $\rm L_{Ly\alpha}$ that we observe represents a lower bound. Taking the bounds of $\rm L_{H_{\beta}}$ that we obtain from Spitzer provides us with a potential range of $0.02 < f_{\rm esc}(\rm{Ly\alpha}) < 0.32$, while the best-fit $\rm L_{H_{\beta}}$ from BAGPIPES returns $f_{\rm esc}(\rm{Ly\alpha}) = 0.09^{+0.07}_{-0.04}$, while taking the bounds on $\rm L_{Ly\alpha}$ the best-fit value ranges from $0.09 < f_{\rm esc}$(LyC) $< 0.18$.

The bounds of the H$\beta$ flux as well as the best-fit value and the associated escape fraction for the stacked spectrum can be found in Table~\ref{tab:fesc}. Using relations from \citet{Maji+22} we can obtain the escape fraction of the Lyman-continuum (LyC) which is found to be lower than the \Lya escape fraction, hence we take the bounds on $f_{\rm esc}$(LyC) to be the same as those on $f_{\rm esc}$(\Lya) (also reported in Table~\ref{tab:fesc}).

\begin{figure*}
	\includegraphics[width=\textwidth]{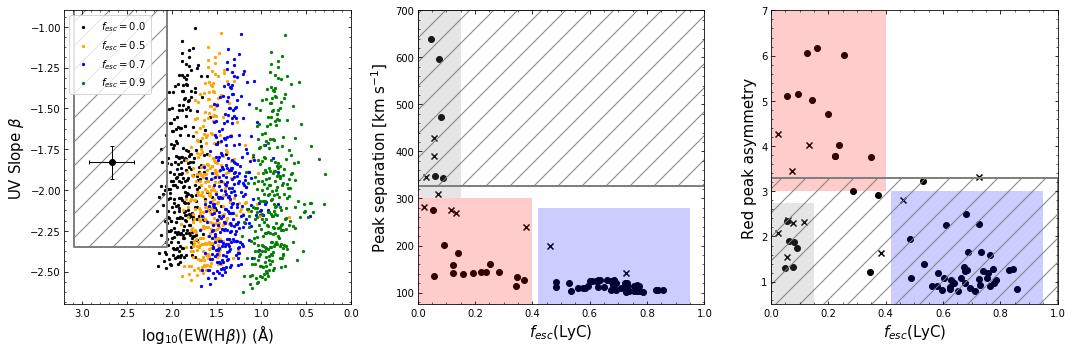}
    \caption{The indirect diagnostics of $f_{\rm esc}$(LyC), with the region which our stack resides indicated with a grey hatched box. (Left) The EW(H$\beta$) of simulated $z = 7-9$ galaxies against their UV slope $\beta$, assuming a Calzetti attenuation law with $E(B - V)_{\rm{stars}} = E(B - V)_{\rm{neb}}$ \citep[from][]{Zackrisson+17}. Their Lyman-continuum escape fraction is denoted by their colour, $f_{\rm{esc}}(\rm{LyC}) = 0.0, 0.5, 0.7, 0.9$ correspond to red, orange, green and blue respectively. The grey hatched region indicates the location of our stack using Spitzer data, while the black data point indicates the position using the BAGPIPIES best-fit on the SED. (Centre) The peak separation of \Lya profiles against their LyC escape fraction \citep[from][]{Kakiichi+21}. The dots denote simulated results from \citet{Kakiichi+21}, while crosses indicate the results for $z \sim 0.3$ LyC-detected galaxies from \citet{Izotov+16,Izotov+18a,Izotov+18b}. The coloured regions indicate the three regimes of LyC escape - leakage by full break, through holes and small leakage with few or no holes indicated by blue, red and grey respectively. (Right) The red peak asymmetry of \Lya profiles against their LyC escape fraction \citep[from][]{Kakiichi+21}. The markers and shading are the same as the central panel.}
    \label{fig:Diagnostics}
\end{figure*}

We additionally consider the relation between the UV slope $\beta$ and the EW(H$\beta$) first determined by \citet{Zackrisson+13} at redshifts $z > 6$ and then at $z \approx 7-9$ by \citet{Zackrisson+17} by studying the evolution in synthetic galaxy spectra with changing $f_{\rm esc}$(LyC). The results of this analysis can be seen in figure~\ref{fig:Diagnostics} where galaxies with log(EW(H$\beta$)) $\gtrsim 2$ exclusively have $f_{\rm esc}$(LyC) $= 0$. As is clear in figure~\ref{fig:Diagnostics} both the best-fit EW(H$\beta$) and UV slope $\beta$ \citep[corrected for dust extinction following][]{Meurer+99} returned from BAGPIPES and the range in EW(H$\beta$) and UV slope $\beta$ \citep[determined following][]{Bouwens+14} that can be estimated from Spitzer data (the values of which are reported in Table~\ref{tab:fesc}) constrain our stack to a region of the figure that is not compatible with an $f_{\rm esc}$(LyC) $\gg 0$. 

We do note that this diagnostic is potentially limited in its ability to diagnose high escape fractions given examples of low redshift galaxies that have high $f_{\rm esc}$(LyC) and high EW(H$\beta$) \citep[eg.][]{Izotov+18b}. The diagnostic requires many assumptions in order to estimate $f_{\rm esc}$(LyC), notably in the stellar models employed. Changing these stellar models can significantly affect the result of the diagnostic, as seen when binary evolution is considered \citep[see figure 6 in][]{Zackrisson+17}. However, we are aware of the main outcome of \citet{Zackrisson+17} - that galaxies with $f_{\rm esc}$(LyC)$>$0.5 should have EW(H$\beta)<30$\,\rm{\AA}. Therefore, we instead consider a high EW(H$\beta$) to be a necessity for low $f_{\rm esc}$(LyC), although perhaps not sufficient. Given the inclusion of multiple diagnostics all indicating low $f_{\rm esc}$(LyC) we consider the potential uncertainty surrounding this diagnostic not to be a significant issue.

We additionally note that \citet{Zackrisson+17} provides this diagnostic for a range of different dust attenuation laws. While the panel in figure~\ref{fig:Diagnostics} assumes a Calzetti attenuation law with $E(B - V)_{\rm{stars}} = E(B - V)_{\rm{neb}}$, we find that the conclusion, that our stack lies within a region of the diagram corresponding to $f_{\rm esc}$(LyC) $= 0$, is consistent regardless of the dust attenuation law used in \citet{Zackrisson+17}.

\subsection{Ly\texorpdfstring{$\alpha$}{TEXT} profile}

Figure~\ref{fig:LyaProfile} clearly indicates that for all of our targets, we find the red-peak of the \Lya profile to be offset from line centre by $\sim 340$ km/s, such a large separation is indicative of a low Lyman-continuum escape fraction \citep{Izotov+18b, Gazagnes+20, Kakiichi+21}. The large offset of the red-peak is additionally present in the stacked spectrum, as well as a clear asymmetry. This asymmetry, A, is the ratio of the blue-to-red flux of the red peak (as defined in \citealt{Kakiichi+21}). We find that the two targets for which we are able to observe the shape of the red-peak profile, we observe clear asymmetry. 

\section{Discussion} \label{sec:Discussion}

The limits placed on $f_{\rm esc}$(\Lya) from photometry discussed in Section~\ref{sec:Photometry Results} are already low enough to rule out the possibility of these three massive, bright galaxies currently being significant contributors to re-ionization. However, we wish to use multiple diagnostics in order to confirm these findings. The results from our \Lya stack are therefore crucial to further constrain the escape fraction. 

\subsection{Interpretation of the velocity offset}
While the velocity offset of \Lya from the systemic redshift initially appears as though it may be primarily driven by outflows in these massive galaxies, we believe this to be unlikely. \citet{Neufeld+90} and \citet{Michel-Dansac+20}, using a static medium with large neutral hydrogen column densities, find velocity offsets in their simulations that are comparable to those that we observe indicating these velocity separations are achievable within simulations without modelling for outflows.

Any such shift in the \Lya profile due to the expansion velocity, $v_{\rm exp}$, of neutral gas would still result in the expected double-peaked profile of \Lya centred on the systemic redshift. Results from \citet{Verhamme+15} indicate that increasing $v_{\rm exp}$ has the effect of reducing the peak separation, such that when $v_{\rm exp} > 300$ km/s, they cannot recreate the red peak to line-centre separation that we observe. These results constrain the neutral gas column density in our stack to be greater than $10^{20} \rm{cm}^{-2}$ and any outflow velocity $v_{\rm exp} < 300$ km/s.

The conclusion that any expansion velocity will act to reduce peak separation thus allows us to conclude that our red-peak offset is a minimum separation. Additionally, most diagnostics of escape fraction that use the separation between the peaks of the \Lya profile are based on observations of lower mass galaxies than our targets and therefore using any such diagnostic is challenging \citep{Izotov+18b, Gazagnes+20}. Instead we choose to use the relation determined by simulations from \citet{Kakiichi+21}, allowing us to avoid mass biases. The observed red-peak-offset in Figure~\ref{fig:LyaProfile} can be used as a lowest bound for the blue-red peak separation given that we know the blue peak lies on the blue side of line-centre. As such, we expect the blue-red peak separation to far exceed the red-peak offset of $\sim 300$ km/s and comparing this to the relation from \citet{Kakiichi+21} we find that $f_{\rm esc}$(\Lya)$ \lesssim 10 \%$ and \Lya photons escape through an optically-thick medium with few or no holes.

\subsection{Interpretation of the red-peak asymmetry}
\begin{figure*}
	\includegraphics[width=\textwidth]{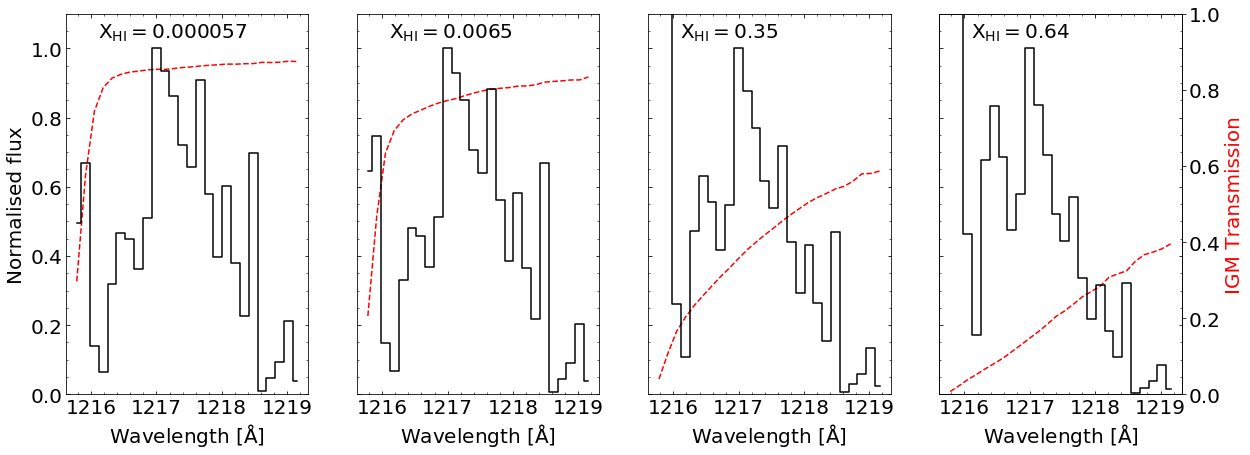}
    \caption{The potential intrinsic stacked spectrum created by dividing the observed spectrum by the IGM transmission for a range of different volumetric neutral fractions indicated in the top left of each panel. The spectrum normalised by the red peak flux is indicated by the solid black line, while the IGM attenuation curve associated with the volumetric neutral fraction, taken from \citet{Garel+21}, is indicated by the red dashed line.}
    \label{fig:IGM}
\end{figure*}
\label{sec:asymmetry}
This asymmetry allows us to quantify the amount \Lya photons have to scatter, in doing so creating a broad wing component of the emission line, in order to escape the galaxy. A high asymmetry (A $> 3$) is hence indicative of \Lya photons having multiple routes to escape and hence scatter significant amounts in order to find low-density channels to escape the galaxy (leakage through holes), while a low asymmetry (A $< 3$) is indicative of \Lya photons only having one method of escape possible either through predominantly optically-thin (leakage by full break) or optically thick (small leakage due to few or no holes) media \citet{Kakiichi+21}. Therefore we can use the asymmetry to attempt to diagnose the properties of the medium through which the \Lya photons have traversed.

The asymmetry that we observe, in Figure~\ref{fig:LyaProfile}, is an upper bound on the asymmetry of the intrinsic spectrum. This is due to IGM attenuation reducing the flux close to line-centre hence reducing the flux between the red-peak and line-centre relative to the flux on the red side of the red-peak, therefore the observed asymmetry is greater than the intrinsic asymmetry. As such we find the asymmetry, A $< 3$, results in the interpretation that \Lya photons have either escaped through a full break environment or by leakage through few or no holes. Given the aforementioned limits on the escape fraction ($f_{\rm esc}$(\Lya)$ \lesssim 10 \%$) we can constrain ourselves to small leakage without the presence of optically-thin channels \citep{Kakiichi+21}.

\subsection{Effects of IGM attenuation} \label{IGM Attenuation}
\begin{figure*}
	\includegraphics[width=\textwidth]{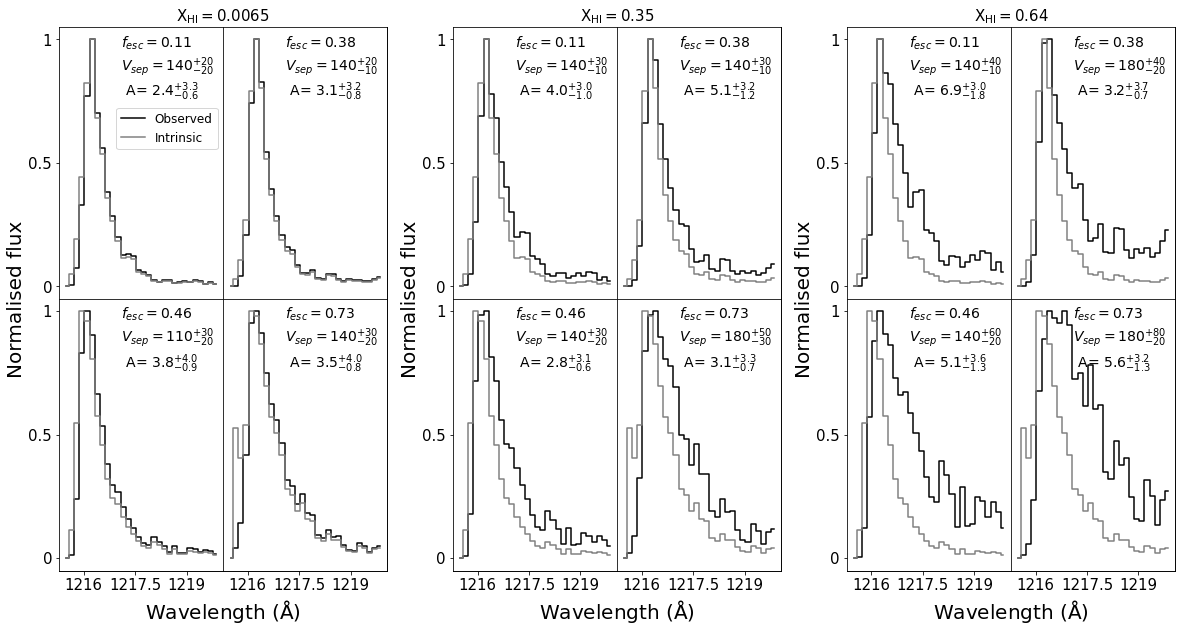}
    \caption{The simulated spectra of high redshift, high $f_{\rm esc}$(LyC) galaxies for varying assumed neutral hydrogen fractions. Each column uses an increasing neutral hydrogen fraction from left to right, that in turn dictates which IGM attenuation curve, taken from \citet{Garel+21}, is applied to the original spectrum. Each panel includes the same 4 galaxies from \citet{Izotov+18a,Izotov+18b} (clockwise from top left sub-panel: J1011+1947, J1256+4509, J1243+4646, J1154+2443), whose Lyman continuum escape fraction are indicated at the top of each sub-panel. The simulated observed spectrum normalised by the red peak flux is indicated by the solid black line, while the original spectrum (with resolution degraded) is indicated in grey.}
    \label{fig:Highfesc}
\end{figure*}
Neutral hydrogen in the IGM causes attenuation of \Lya close to line-centre \citep[see][]{Garel+21}, therefore in order to confirm that we do not misidentify the location of the red-peak flux, we divide our observed stacked spectrum through by attenuation curves for varying volumetric neutral fraction, taken from \citet{Garel+21}. Given our target galaxies all likely reside within large ionized bubbles we do not expect significant IGM absorption. Figure~\ref{fig:IGM} indicates the effect of correcting for the different IGM transmission curves and we see no notable difference in our observed spectra even at the most extreme volumetric neutral fraction, indicating our observed red peak separation likely trace that intrinsic to the galaxy before IGM absorption of \Lya. While we do see a notable decrease in the asymmetry, as predicted in Section~\ref{sec:asymmetry}, we have already consider the asymmetry to be a lower bound and as such this does not affect the interpretation of the result. 

We do note a significant increase in the flux at line-centre, in Figure~\ref{fig:IGM}, due to the effectively zero transmission through the IGM at that wavelength. This is merely an artefact of noise being divided through by a number tending to zero rather than any physical intrinsic property of the galaxy. We know this to be true as \Lya emission will immediately be absorbed at line-centre by any neutral hydrogen within the host galaxy and as such we must observe negligible flux at line-centre escaping the host galaxy.

Finally, in order to confirm that we are not being affected by high escape fraction interlopers that due to IGM transmission, spectral resolution and noise are being interpreted as having a low escape fraction, we attempt to recreate high redshift observations of galaxies with $f_{\rm esc}$(LyC) greater than our sample ($f_{\rm esc}$(LyC)$>0.1$). We use the spectra of galaxies from \citet{Izotov+18a,Izotov+18b} as examples of \Lya profiles associated with an $f_{\rm esc}$(LyC) greater than our sample up to a value of 72\% at low redshifts ($z \sim 0.3$). We apply the IGM attenuation curves from \citet{Garel+21} to the \Lya profile, we then reduce the resolution of these spectra down to the resolution of our stacked \Lya profile and finally we use the MC error propagation described in Section~\ref{sec:Method} to estimate uncertainties on the V$_{\rm red}$ and asymmetry of each galaxy given a noise level similar to our stacked spectrum (by assuming the peak flux to be at SN = 5). 

Figure~\ref{fig:Highfesc} shows our simulated high redshift, high $f_{\rm esc}$(LyC) spectra with various IGM transmission curves applied for differing neutral hydrogen fractions. We find that applying the IGM transmission and degrading the resolution of the spectra result in velocity offsets that are consistent with the original spectra, even for the most extreme neutral hydrogen density, to within 40 km/s. This peak separation when considered in the context of the \citet{Kakiichi+21} diagnostic appears to indicate these galaxies are likely high $f_{\rm esc}$(LyC). All of these galaxies exhibit red peak offsets $\ll 300$ km/s thus allowing us to conclude that the observation of a peak offset of $\sim 300$ km/s is not only indicative of a low $f_{\rm esc}$(LyC) but also that we are likely not being affected by high $f_{\rm esc}$(LyC) interlopers. We do however note that the asymmetry is more challenging to understand as a function of the neutral fraction. It is clear that for a sharp \Lya red-peak, increasing the neutral density will act to increase the flux on the red side of this peak hence increasing the asymmetry. When the red-peak is more broad, increasing the neutral density can push the peak of the \Lya profile red-ward and hence act to increase the amount of flux on the blue side of the peak. This complicated interplay of effects leads to a highly uncertain asymmetry in some of our \Lya profiles. Therefore, the use of asymmetry to diagnose $f_{\rm esc}$(LyC) alone at high redshifts, where neutral hydrogen in the IGM causes large uncertainties on the intrinsic asymmetry, should be avoided. However, the asymmetry of our stacked spectrum appears relatively well defined with a sharp drop in flux blue-ward of the peak flux. Therefore, we conclude that our asymmetry is most likely an upper-bound asymmetry, where the relative boosting of flux to the red side of the \Lya peak due to IGM attenuation, is the most likely of the two aforementioned effects at play. Given the relatively small uncertainty on the intrinsic \Lya asymmetry of our stacked spectrum we believe that, in combination with multiple other diagnostics, the observed asymmetry can be used to infer the ability of \Lya photons to escape their host galaxy, with the caveat that this diagnostic should not be used on high redshift LAEs without the use of other supplementary diagnostics.

\section{Summary} \label{sec:Summary}

In order to probe the potential ionising properties of the most massive (log$_{10}\rm(M_{\star} [M_{\odot}]) > 10$), UV-bright ($M_{UV} \sim -22$), high redshift ($z > 7$) galaxies, we target all archival data on telescopes with resolution $R \gtrsim 3000 $, allowing us to obtain a resolved \Lya profile. We find a total of four observations of three satisfactory galaxies with \Lya emission, totalling an exposure of 24 hours and 45 minutes. Using a median stacking method we obtain a deep stacked spectrum representing massive, UV-bright, high redshift \Lya leaking galaxies. Through the analysis of the stacked \Lya profile, using the red-peak velocity offset from line-centre and the red-peak asymmetry, we deduce the Lyman-continuum escape fraction to be less than 10$\%$ and that the few Lyman-continuum photons that do escape, escape through an optically-thick medium with few or no holes.  Through the use of Spitzer observations of our target galaxies, stacking these and SED-fitting using BAGPIPES we obtain bounds on the recombination line $\rm{H}\beta$. Given this we constrain the escape fraction to $9 \% < f_{\rm esc}$(\Lya)$ < 18 \%$ in strong agreement with the results of our stacked \Lya profile. We additionally confirm that neither IGM attenuation or a significant outflow velocity could affect our conclusion regarding a low $f_{\rm esc}$(\Lya) for massive, UV-bright, high redshift galaxies. Our study shows that despite the fact the 3 galaxies analysed lie within ionised bubbles, they are not capable themselves of ionising their own bubbles. 

However, we emphasize that our result is obtained using only 4 datasets of 3 different galaxies at $z\geq$7 -- the only observations currently available in telescope archives. Increasing the number of \Lya detections at $z\geq$7 with high-resolution spectrographs is therefore crucial to confirm our conclusions. Furthermore, the high-fraction of neutral gas underlying galaxies within the epoch of reionisation limits the detection of \Lya to galaxies in overdense regions. Spectrographs with a large field-of-view will therefore be ideal instruments to push forward this project. MOONS, a 3$^{rd}$ generation instrument at the Very Large Telescope, will be one of those. It combines high-resolution (R$>$4000), a large field of view ($\sim$500 arcmin$^2$) and a huge number of fibres ($\sim$1000). 

\section{Acknowledgements}
We thank the anonymous referee for providing helpful comments which improved
the quality of this paper.
CW and NL acknowledge advice and comments from Debora Sijacki, Martin Haehnelt, Roberto Maiolino, Sergio Martin-Alvarez and Yuxuan Yuan that helped to direct our analysis and the diagnostics used. 
CW acknowledges support from the Science and Technology Facilities Council (STFC) for a Ph.D. studentship.
NL acknowledges support from the Kavli foundation. 
 This research has made use of the Keck Observatory Archive (KOA), which is operated by the W. M. Keck Observatory and the NASA Exoplanet Science Institute (NExScI), under contract with the National Aeronautics and Space Administration. Based on observations collected at the European Southern Observatory under ESO programme 097.A-0043(A). 
This work is based on observations taken by the 3D-HST Treasury Program (GO 12177 and 12328) with the NASA/ESA HST, which is operated by the Association of Universities for Research in Astronomy, Inc., under NASA contract NAS5-26555
 
\section{Data Availability}
The data underlying this article will be shared on reasonable request to the corresponding author.

\bibliography{references}{}
\bibliographystyle{aasjournal}

\end{document}